\documentclass[aps,prl,floats,twocolumn,showpacs,superscriptaddress]{revtex4-1}

\usepackage{graphicx}
\usepackage{amssymb}
\usepackage{color}

\graphicspath{{c:/fortran/aston/}}

\begin{document}

\title{Unveiling temporal correlations characteristic to phase transition in the intensity of a fibre laser radiation}

\author{A. Aragoneses}
\affiliation{Departament de F\'{\i}sica i Enginyeria Nuclear, Universitat Polit\`ecnica de Catalunya, 08222 Terrassa, Spain}
\affiliation{Duke University, Physics Department, Box 90305, Durham, North Carolina 27708, USA}

\author{L. Carpi}
\affiliation{Departament de F\'{\i}sica i Enginyeria Nuclear, Universitat Polit\`ecnica de Catalunya, 08222 Terrassa, Spain}

\author{N. Tarasov}
\affiliation{Aston Institute of Photonic Technologies, Aston University, UK}
\affiliation{Institute for Computational Technologies, SB RAS, Novosibirsk 630090, Russia}

\author{D. V. Churkin}
\affiliation{Aston Institute of Photonic Technologies, Aston University, UK}
\affiliation{Institute of Automation and Electrometry SB RAS, Novosibirsk, Russia}
\affiliation{Novosibirsk State University, Russia}

\author{M. C. Torrent}
\affiliation{Departament de F\'{\i}sica i Enginyeria Nuclear, Universitat Polit\`ecnica de Catalunya, 08222 Terrassa, Spain}

\author{C. Masoller}
\affiliation{Departament de F\'{\i}sica i Enginyeria Nuclear, Universitat Polit\`ecnica de Catalunya, 08222 Terrassa, Spain}

\author{S. K. Turitsyn}
\affiliation{Aston Institute of Photonic Technologies, Aston University, UK}

\begin{abstract}
We use advanced statistical tools of time-series analysis to characterize the dynamical complexity of the transition to optical wave turbulence in a fibre laser. Ordinal analysis and the horizontal visibility graph applied to the experimentally measured laser output intensity reveal the presence of temporal correlations during the transition from the laminar to the turbulent lasing regimes. Both methods unveil coherent structures with well defined time-scales and strong correlations both, in the timing of the laser pulses and in their peak intensities. Our approach is generic and may be used in other complex systems that undergo similar transitions involving the generation of extreme fluctuations.
\pacs{05.45.Tp; 89.75.-k; 89.70.Cf; 89.75.Hc} 
\end{abstract}

\maketitle

Fibre lasers are important practical devices that represent complex nonlinear systems with many degrees of freedom \cite{cid,sergei_prl, FL5}.  Typically, the output of a fibre laser involves nonlinear interactions of millions of longitudinal cavity modes in regimes far from thermal equilibrium \cite{dima_oe}.  
In general, wave dynamics in fibre lasers is highly complex, as in other optical wave turbulence systems \cite{OT,WT1}. Though the underlying physical effect, nonlinear four-wave mixing, is purely deterministic and well understood, it is also well known that a deterministic dynamical description is not fully adequate in this problem, and statistical tools such as entropy and complexity measures should be used to characterize the complex fluctuations in the generated output signals. Within this framework of wave turbulence, the role of ``temperature" is played by optical noise that occurs in the gain medium, which in fibre lasers leads to an effective ``nonlinear noise" due to four-wave-mixing.

Recently, the analogy between hydrodynamic transition to turbulence and change of operational regimes in fibre lasers has been studied both experimentally and theoretically \cite{FLOT1}. A transition from highly ordered lasing regime to more irregular lasing, characterized by extreme, apparently random intensity fluctuations was reported. Such transition, being a relevant example of a phase transition in a one dimensional physical system, was shown to be accompanied by the occurrence of coherent spatio-temporal structures \cite{FLOT1}. 

In this work we address an important question relevant to practical identification of such structures: are there underlying correlations and/or specific time-scales in the easily measurable intensity fluctuations of laser radiation? In order to investigate this issue we use two nonlinear analysis tools: ordinal analysis \cite {bandt_PRL_2002} and the horizontal visibility graph \cite{luque_2009}. We show that both methods allow to clearly identify the presence of long-range temporal correlations in the experimentally measured laser output intensity.


In our experiments, we measure an output temporal intensity dynamics of a quasi-CW Raman fiber laser formed of  1 km of normal dispersion fiber placed between two fiber Bragg gratings acting as cavity mirrors \cite{FLOT1}. State-of-the-art experimental capabilities allowed us to register extremely long time traces with total number of intensity data points of 50 million. Taking into account the discretisation time of 12.5 ps, the intensity dynamics over  625 $\mu$s could be captured. In order to be able to compare among time-series recorded at different pump power, each time-series is normalized to have zero-mean and unit variance. Depending on power, the generation regime can be considered as "laminar" or "turbulent" \cite{FLOT1}. The transition occurs at pump power 0.9 W (see \cite{FLOT1} for details). Despite the radically different coherence properties of radiation in these two regimes, the output intensity, $I(t)$, looks similar and irregular at all powers, as seen in Fig. 1(a), with typical intensity probability distribution function (pdf) of intensity values, $p(I(t))$, shown on Fig. 1(b).

\begin{figure}
   \includegraphics[width=0.48\columnwidth]{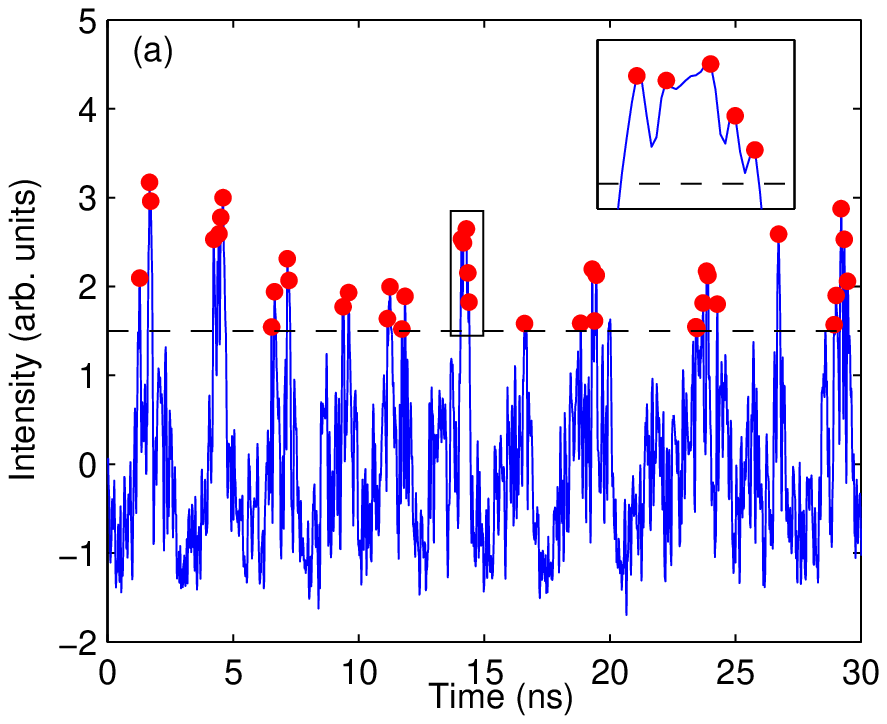}
   \includegraphics[width=0.48\columnwidth]{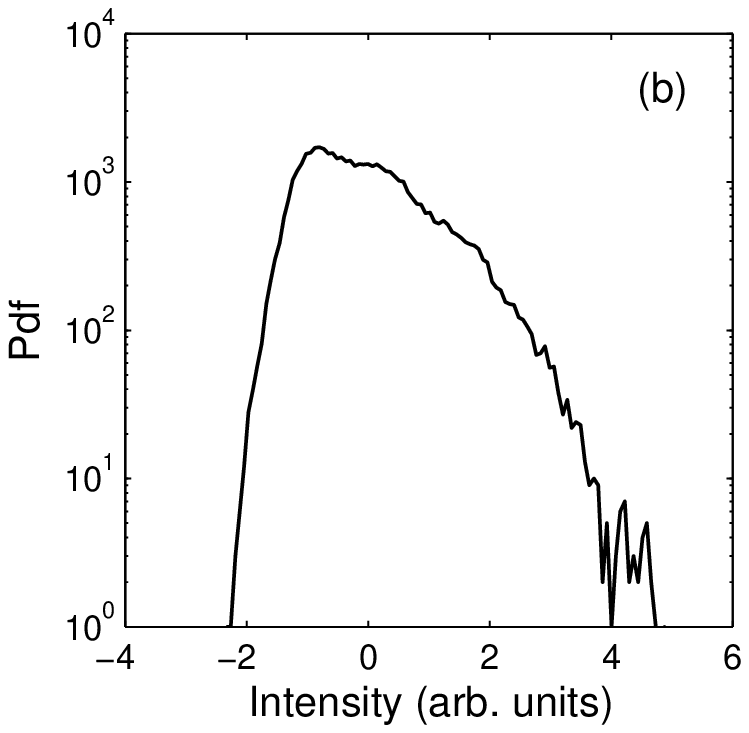}
   \includegraphics[width=0.48\columnwidth]{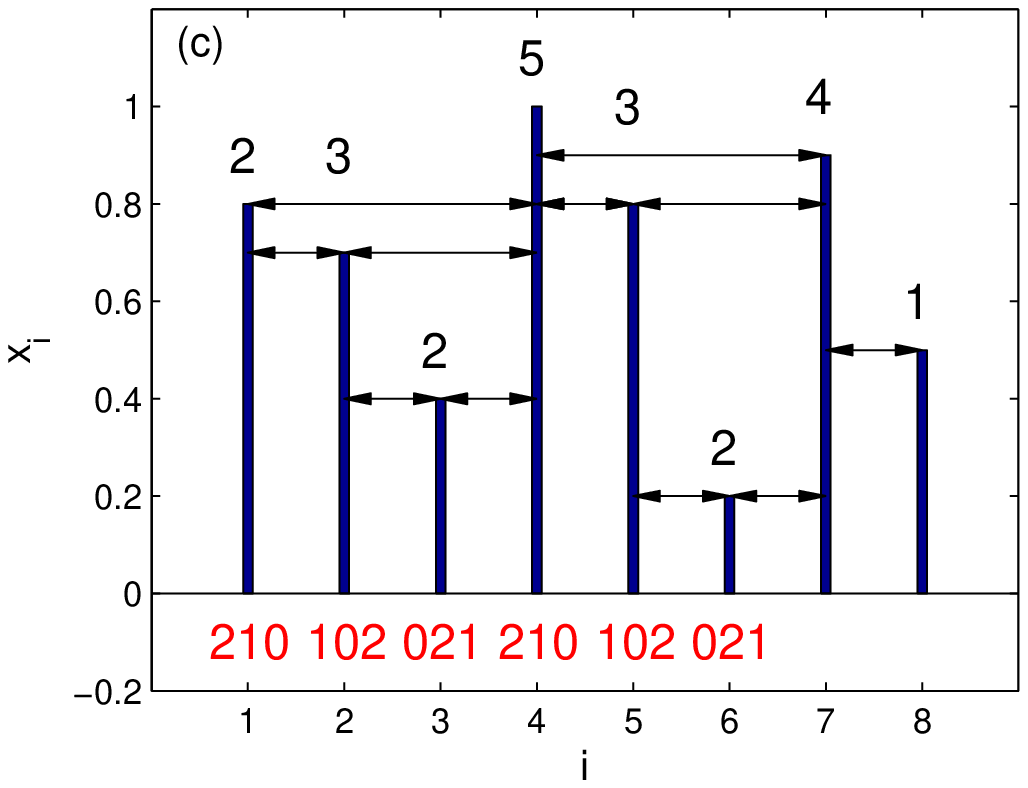}
  \includegraphics[width=0.48\columnwidth]{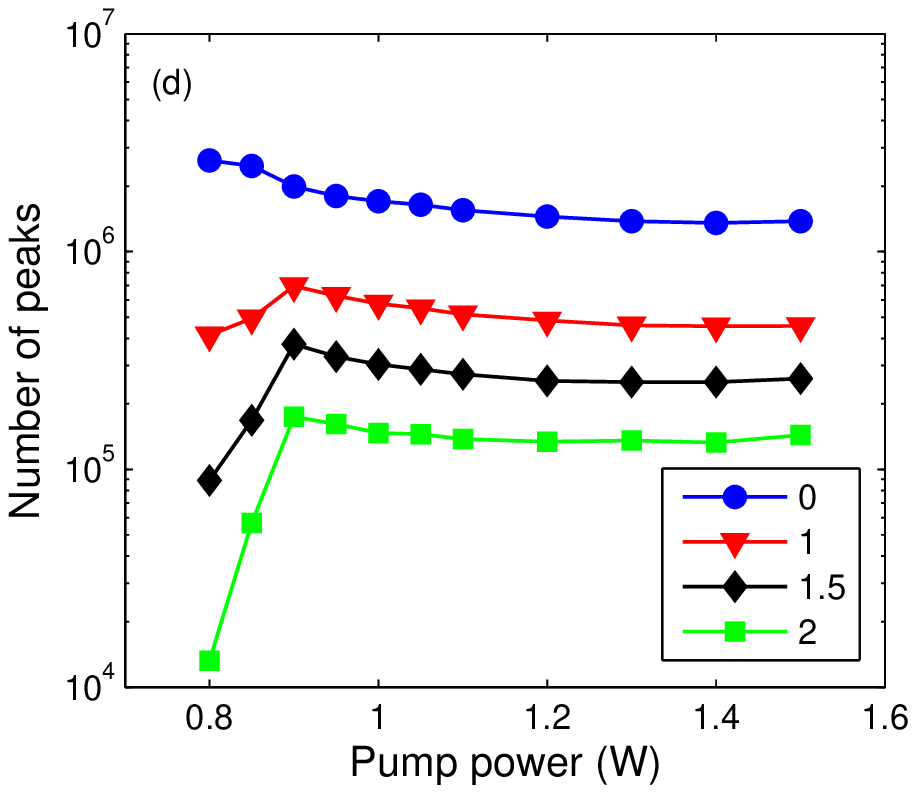}
  \caption{(a) Stochastic dynamics of a quasi-CW Raman fibre laser: intensity time-series measured at pump power 0.9 W (the dots indicate the peak values above a threshold, indicated with a dashed line, and the inset displays a detail). (b) Pdf of the intensity values. (c) Schematic representation of the two analysis methods: the values $\{x_i\}$ in a time series are represented with vertical bars, the ordinal patterns formed by $(x_i, x_{i+1}, x_{i+2})$ are indicated in red, the links in horizontal visibility graph are indicated with arrows, and the numbers indicate the degree (the number of links) of each data point (graph node). (d) Number of intensity peaks vs the pump power for various thresholds (threshold values are shown by different colours).}
  \label{fig:fig1}
\end{figure}

As a starting step of our analysis, we  generate two new data sets from experimentally measured intensity dynamics, $I(t)$. In particular, we filter out  from the initial intensity dynamics only local intensity peaks and create a sequence of intensity peak heights, $\{I_{\max,i}\}$, which are above a certain threshold, also shown in Fig. 1(a). Because each time series is normalized to zero mean and unit variance, the thresholds used are in units of the standard deviation, $\sigma$: with threshold 0, the peaks considered are only those above the mean value; with threshold 1, the  peaks are only those above $\sigma$. The number of the peaks found in intensity dynamics measured at different pump power level is shown on Fig. 1(d) depending on the threshold value. We note that, if the threshold is low, the number of peaks decreases with the pump power, but if the threshold is high, the number of peaks increases with the pump power until 0.9 W, where it is maximum, then, with further increase of the pump power, the number of peaks diminishes. The transition between two different generation regimes --- laminar and turbulent regime --- could be easily detected at 0.9 W. In the following we fix the threshold equal to 2, i.e., we analyze only the intensity peaks that are above $2\sigma$. As each local intensity peak has also a time instant $T_i$ at which it occurs, we generate in parallel another data set: a  sequence of time intervals between local intensity peaks, $\{\Delta T_i\}$.

To investigate the complexity of the dynamics and to uncover hidden temporal correlations we use two methods of time-series analysis, which are represented schematically in Fig. 1(c). The first one, known as \emph{ordinal analysis} \cite{bandt_PRL_2002}, transforms a time series $\{x_i\}$ (where $\{x_i\}$ is either the sequence of peak heights, $\{I_{\max,i}\}$, or the sequence of time intervals between peaks, $\{\Delta T_i\}$) into a sequence of symbols (referred to as ordinal patterns, OPs), by considering the order relation among $D$ values of the time-series. For example, there are 2 different ordinary patterns of length 2: pattern `01' if $x_i<x_{i+1}$ and pattern `10' if $x_i>x_{i+1}$. If $D=3$, there are 6 possible patterns: $x_i<x_{i+1}<x_{i+2}$ gives `012', $x_{i+2}<x_{i+1}<x_{i}$ gives `210', etc. The number of possible patterns of the given length $D$ is $D!$. In this way, a sequence of patterns could be generated from the sequence of the peak heights or from the sequence of time intervals between the peaks.

After defining the sequence of patterns, one can calculate the probability to find the given pattern in the data set, $p_i$. The entropy computed from their probabilities, $p_i$, of occurrence in the time series, $S_{PE} = - \sum p_i \log p_i$, known as permutation entropy, has been shown to be an appropriated measure of the complexity of a time-series \cite{bandt_PRL_2002,zanin,special_issue}. When there are no serial correlations in the time-series $\{x_i\}$, then all the patterns are equally probable and $S_{PE}\sim \log D!$. On the contrary, when there are serial correlations, then the OPs are not all equally probable, and the permutation entropy will be $S_{PE}<\log D!$. In the following we refer to the normalized entropy, $S_{PE} /\log D!$, as PE entropy. Thus, with an appropriate choice of pattern length $D$, the OP probabilities and the PE entropy will capture the existence of underlying correlations in the time series.

To verify independently presence of correlations, we also apply the so-called \emph{horizontal visibility graph} (HVG) method~\cite{luque_2009}. In this approach a time series, $\{x_i\}$, is converted into a graph by considering each data point, $x_i$, as a node. Any two nodes are connected by a link (or edge) if horizontal visibility exists between them [see Fig. 1(c)]: $x_i$ and $x_j$ are connected if it is possible to trace a horizontal line linking $x_i$ and $x_j$ not intersecting intermediate data; mathematically, this means that $x_i$ and $x_j$ are connected if: $x_i,~x_j~> x_n$ for all $i < n <  j$. Note that this graph representation of the time series $\{x_i\}$ takes into account both, the order and the values of the data points. Time series with different dynamics are mapped into graphs that exhibit distinct topological structures \cite{lacasa_pre_2010}. The topology of a graph is characterized by the degree distribution, $p(k)$, that is the probability that a node has $k$ links. Thus, the entropy of the degree distribution, $S_{HVG} = -\sum p_k \log p_k$ (in the following, referred to as HVG entropy), is another measure of the complexity of the time series $\{x_i\}$ ~\cite{ravetti_2014}. The HVG method also allows to analyze different time-scales by constructing the graph not from all the ``raw'' data points, but from lagged data: $\{x_i, x_{i+\tau}, x_{i+2\tau},\dots \}$.

While both analysis methods share the common feature of transforming the time series $\{x_i\}$ into a sequence of integer numbers, $\{k_i\}$ (in the OP case, $k_i\in[1,D!]$ is the pattern label: if `01', $k_i=1$; if `10', $k_i=2$, etc.; in the HVG case, $k_i\in[1,N-1]$ is the degree of $x_i$, with $N$ being the number of data points in the time series), they have important differences: while the OP method requires the pre-definition of the length of the pattern $D$, and does not take into account the values of the data points, the horizontal visibility graph method does not require to pre-define an analysis length, and considers both, the order relation and the actual values of the data points.


\begin{figure}[tbh]
   \includegraphics[width=1.0\columnwidth]{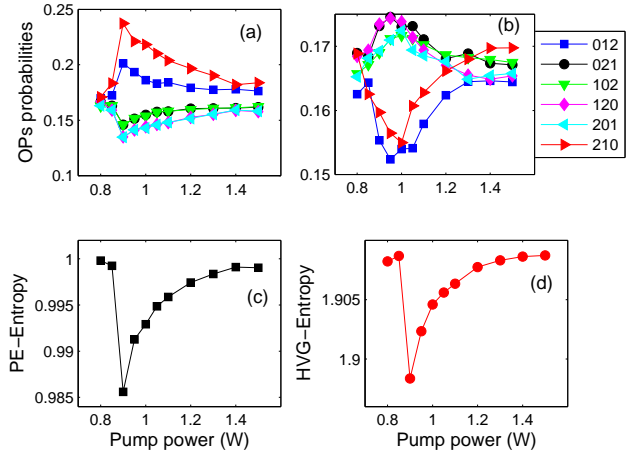}
  \caption{Probabilities ordinal patterns of length $D=3$ vs the pump power calculated  from the sequence of (a) intensity peaks and (b) time intervals between consecutive peaks. Permutation entropy (c) and HVG-entropy (d) calculated  from the sequence of intensity peaks.}
  \label{fig:fig2}
\end{figure}

 Figures 2(a) and 2(b) display the probabilities of the six patterns of length 3 calculated from the sequence of intensity peaks and from the sequence of the time intervals, respectively. We observe that the variation of  the probabilities with the pump power captures the transition between two dynamical regimes: below the transition the OPs are equally probable, while during the transition from laminar to turbulent regime their probabilities are different from equiprobability. It can also be noticed that, during the transition, the patterns 012 and 210 become more probable (less probable) when the OPs are calculated from the $\{I_{\max,i}\}$ sequence (from the $\{\Delta T_i\}$ sequence). We also note that the patterns calculated from the intensity peaks capture more determinism than those computed from the time intervals [notice the difference in the vertical scales of Figs. 2(a) and 2(b)]. This indicates that the timing of the high intensity peaks is more random than their peak values.

 The PE entropy, Fig. 2(c), quantifies this effect by decreasing sharply at the transition power. A similar behavior is observed when computing the HVG-entropy, Fig. 2(d). We have verified that, during the transition, the OP probabilities (both, for the intensity peaks and for the time intervals) are not consistent with the null hypothesis of full stochasticity: the probabilities lie outside the region consistent with random OPs, $p\pm 3 \sigma_p$, where $p=1/D!$ and $\sigma_p = \sqrt{p(1-p)/N}$ with $N$ being the number of data points \cite{aragoneses_2014}. In Fig. 2(c) the PE-entropy was computed from $D=3$ OPs; a similar plot was obtained with $D=4$ and $D=5$ OPs (not shown). Larger $D$ values were not considered due to the finite length of the dataset: while the ``raw'' intensity time series contains 50 million data points, the number of high intensity peaks (above $2\sigma$) shown in Fig. 1(d) is about $10^5$, depending on the pump power.

\begin{figure}[tbh]
\includegraphics[width=1.0\columnwidth]{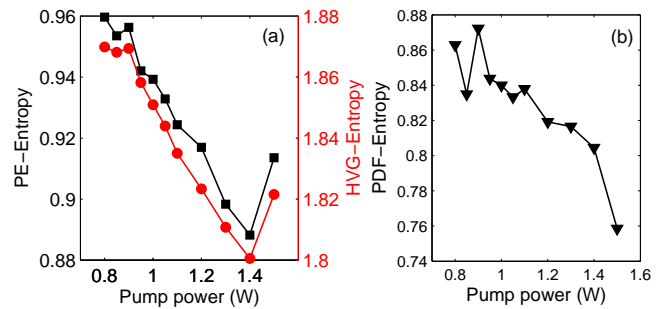}
  \caption{(a) The permutation entropy (left vertical axis) and the HVG-entropy (right vertical axis) vs the pump power. Calculations are performed for the ordinal patterns of length $D=3$. (b) Pdf-entropy calculated from the distribution of intensity values.}
  \label{fig:fig3}
\end{figure}
The probabilities of patterns 012 and 210 provide a measure of the persistence of the time-series, i.e., the probability that the sign of $x_{i}-x_{i-1}$ persists in the next step \cite{bandt_2007}. Thus, at the transition, if there are two consecutive peaks with increasing height, the next peak is likely to be larger than the previous one (and if there are two consecutive peaks with decreasing height, the next one is likely to be smaller than the previous one); on the contrary, in the sequence of time-intervals, two consecutive intervals that are increasingly long ($\Delta T_i<\Delta T_{i+1}$) are likely to be followed by shorter interval ($\Delta T_{i+1}>\Delta T_{i+2}$), and two consecutive decreasing intervals ($\Delta T_i>\Delta T_{i+1}$) are likely to be followed by a longer one ($\Delta T_{i+1}<\Delta T_{i+2}$).


\begin{figure}[tbh]
  \includegraphics[width=1\columnwidth]{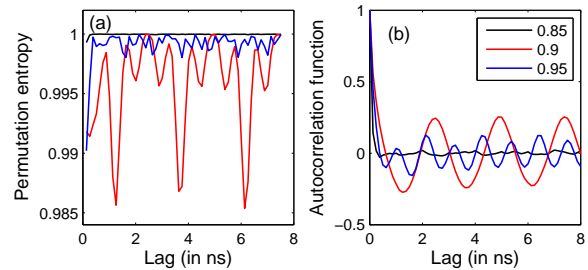}
  \caption{(a) Permutation entropy vs the lag-time before (0.85 W), at (0.9 W), and after (0.95 W) the transition to optical turbulence. (b) Autocorrelation function vs lag, for the same pump powers as panel (a).
  }
  \label{fig:fig4}
\end{figure}

Let us note that the optimal value of pattern length $D$ depends of the  length of correlations embedded into the time series \cite{bandt_PRL_2002}. 
To investigate the presence of specific time-scales in the dynamics, we analyze the lagged time series, i.e. the sequence of $\{I_i, I_{i+\tau}, I_{i+2\tau}, \dots \}$, where $\tau$ is the lag time. We begin by considering the case $\tau=1$ (i.e., we analyze all data points). Figure 3(a) displays the PE entropy and the HVG entropy vs the pump power, and same behavior is seen in both entropies: there is a clear transition at pump power 0.9 W, where both entropies smoothly decrease. It is also observed that for the highest pump power, both entropies increase again. This reveals that during the transition there is an increase in the ``ordering'' of consecutive intensity values (that is captured by both entropies, which decrease), but for the highest pump power the trend reverses and the disorder increases. In contrast, the entropy computed in the conventional way and referred as pdf-entropy (i.e. the entropy calculated from the intensity pdfs of the initial intensity dynamics, I(t)) does not capture this behavior: as it can be seen in  Fig. 3 (b), after the transition the pdf-entropy monotonously decreases with the pump power.
 Thus, $S_{PE}$ and $S_{HVG}$ provide consistent information, which complements that gained from the standard pdf-entropy. The good agreement between the PE and HVG entropies, also seen in Figs. 2(c) and 2(d), is remarkable because the two methods transform a time series into a sequence of integer numbers by using very different encoding rules.

By varying the value of the lag time $\tau$, i.e. by taking into account not all points in data sets, but every second ($\tau=2$) point, every third ($\tau=3$) point etc., we are able to identify a specific oscillation time-scale in the intensity time-series during the transition. The PE entropy vs $\tau$ for pump powers below (0.85 W), at (0.90 W) and above (0.95 W) the transition is displayed in Fig. 4(a). Here we can notice that, at the transition, there are specific lags for which the PE entropy decreases sharply. Similar results were obtained with the HVG entropy.

The sharp minima indicate that, for pump power 0.90 W and for specific lags, 6 different patterns of length $D=3$ are not all equally probable, and thus, there are serial correlations in the sequence of lagged intensity values. To explore the length of such correlations, we computed the PE entropy using longer ordinary patterns ($D=4$ and $D=6$) and found that the minima were more pronounced, revealing the existence of long serial correlations. These correlations are not captured by the classical autocorrelation function shown in Fig. 4(b) for comparison purposes, that displays only a smooth variation with $\tau$.

\begin{figure}[tbh]
 \includegraphics[width=0.8\columnwidth]{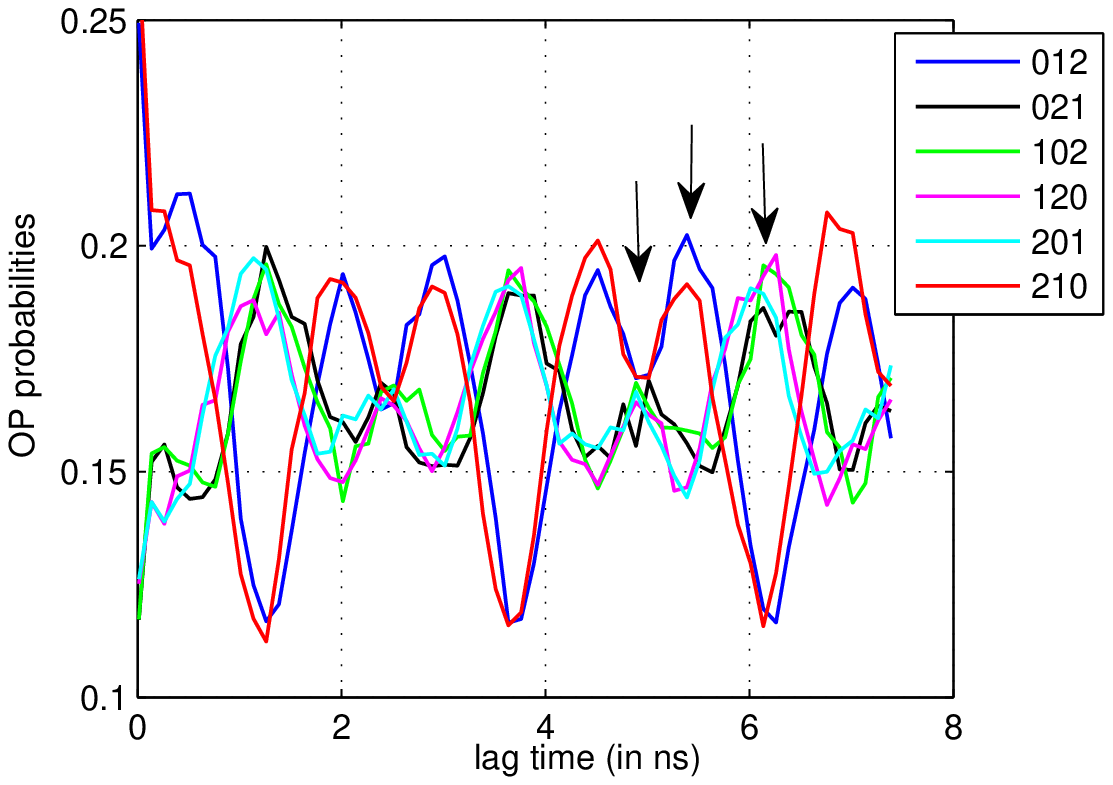}\\
  \includegraphics[width=1\columnwidth]{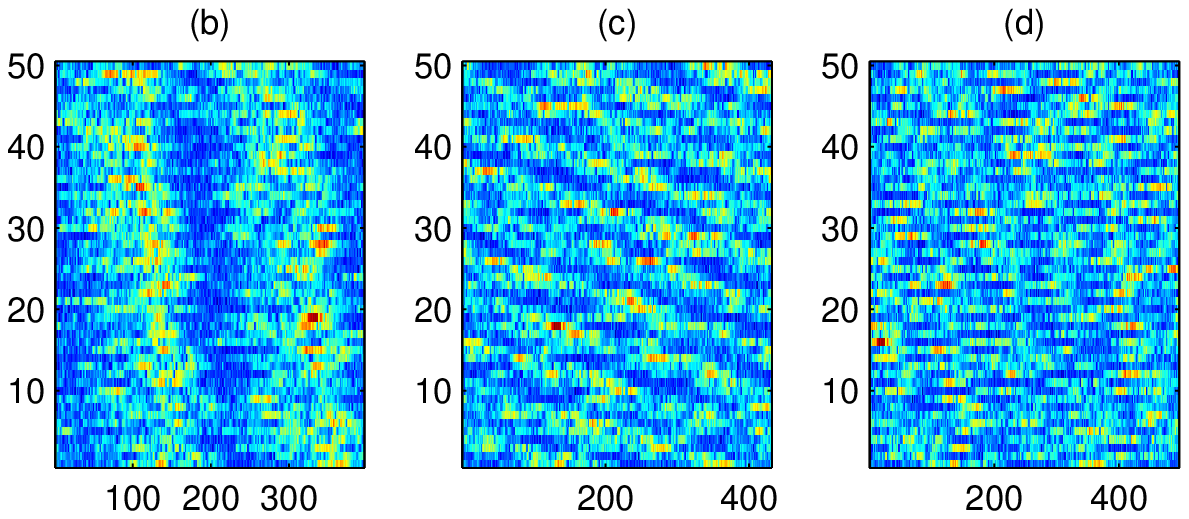}
  \caption{(a) Probabilities of the ordinary patterns of length $D=3$ vs lag time $\tau$. (b)-(d) Spatio-temporal structures identified with the specific lags indicated with arrows in panel (a). The color scale indicates the value of $I_{i}$ with $i=n\tau +j$ and $\tau$=396, 431 and 496 in units of the sampling time. The pump power is 0.9W.}
  \label{fig:fig5}
\end{figure}

To investigate the nature of these correlations we plot on Fig. 5(a) how the probabilities of 6 different ordinary patterns of length $D=3$ depend on the lag time. We note a periodic alternation in which 012 and 210 became the more probable or the less probable patterns. The probabilities of the other four patterns are similar (no clear clusters are seen). The lag values for which 012 and 210 are less probable correspond, as expected, to the lag values where the autocorrelation function is minimum (and negative). However, unexpectedly, the lag values for which 012 and 210 are more probable, do not correspond to the maxima of the autocorrelation; and moreover, for the lag values where the autocorrelation is maxima, the all six ordinary patterns have similar probabilities. These observations suggest that ordinal analysis identifies subtle correlations in the \emph{ordering} of data points, which are not seen by the standard autocorrelation function, that measures correlations in the \emph{values} of the data points.

These ``order correlations'' result into different types of spatio-temporal patterns. Let us recall that the initial intensity spatio-temporal dynamics could be seen as intensity spatio-temporal dynamics \cite{natcomm}. Similar concept has been recently employed for studying interaction of cavity solitons \cite{jang2013ultraweak} and topological solitons as addressable bits \cite{garbin2015topological}. Here  we apply this concept and choose specific lag times defined from Fig.5(a) (shown by arrows) to be used as round-trip times in processing of the initial data series (see \cite{natcomm} for details of plotting spatio-temporal dynamics). Figures 5(b)-(d) display examples with lags such that patterns 012 and 210 are as probable [Fig. 5(b)], more probable [Fig. 5(c)]) and less probable [Fig. 5(d)] than the other four patterns. We can see that these spatio-temporal dynamics of patterns display clear and different coherent structures. These observations can be useful for confronting the predictions of state-of-the-art laser models with empirical data, and the theoretical studies could provide insight into the physical mechanisms underlying these correlations.

To summarize, by applying two independent tools of nonlinear time-series analysis we have uncovered long-range temporal correlations in the intensity output of a fibre laser during the transition to a wave turbulence regime. Output of laser radiation is easily measurable, making these easily implementable methods useful and valuable techniques for investigating coherent structures in complex laser radiation. Both approaches can be applied to any high-dimensional complex systems that undergo similar transitions accompanied by the generation of extreme fluctuations.

\section{Acknowledgements}
This work has been supported by the ERC  project ULTRALASER, the Ministry of Education and Science of the Russian Federation (agreement No. 14.B25.31.0003), the Spanish MINECO (FIS2012-37655-C02-01), EOARD (Grant FA9550-14-1-0359), CNPq Brazil, Russian Foundation for Basic Research (grant 15-02-07925), Presidential Grant for Young researchers (grant 14.120.14.228-MK). N.T. is supported by the Russian Science Foundation (Grant No. 14-21-00110).




\begin{thebibliography}{99}
\bibitem{cid} C. J. S. de Matos, et al.,
Phys. Rev. Lett. \textbf{99}, 153903 (2007).

\bibitem{sergei_prl} S. K. Turitsyn, J. D. Ania-Castanon, S. A. Babin, et al.,
Phys. Rev. Lett. {\bf 103} 133901 (2009).

\bibitem{FL5} S. K. Turitsyn, S. A. Babin, D. V. Churkin, I. D. Vatnik, M. Nikulin, E. V. Podivilov,
Phys. Rep. {\bf542}, 133 (2014).

\bibitem{dima_oe}
D. V. Churkin, A. E. El-Taher, I. D. Vatnik, et al, Opt. Express {\bf 20} 11178 (2012).









\bibitem{OT} A. Picozzi, J. Garnier, T. Hansson, P. Suret, S. Randoux, G. Millot, and D. N. Christodoulides,
Phys. Rep. {\bf542}, 1132 (2014).

\bibitem{WT1} Advances In Wave Turbulence (World Scientific Series on Nonlinear Science: Series A) Hardcover  10 May 2013
Edited by Victor Shrira and Sergei Nazarenko.

\bibitem{FLOT1} E. G. Turitsyna, S. V. Smirnov,	S. Sugavanam,	N. Tarasov,	X. Shu,	S. A. Babin,	E. V. Podivilov, D. V. Churkin,	G. E. Falkovich and S. K. Turitsyn
Nat. Phot. {\bf7}, 783 (2013).



\bibitem{optical_links} A. El-Taher, O. Kotlicki, P. Harper, et al., Laser \& Phot. Rev. {\bf 8} 436 (2014).

\bibitem{bandt_PRL_2002} C. Bandt and B. Pompe,
Phys. Rev. Lett. {\bf 88}, 174102 (2002).

\bibitem{luque_2009}  B. Luque, L. Lacasa, F. Ballesteros and J. Luque,
Phys. Rev. E  {\bf 80}, 046103 (2009).


\bibitem{zanin} M. Zanin, et. al., Entropy \textbf{14}, 1553 (2012).

\bibitem{special_issue} J. M. Amigo, et. al., Eur. Phys. J. Spec. Top. \textbf{222}, 2 (2013).

\bibitem{lacasa_pre_2010} L. Lacasa and R. Toral, 
Phys. Rev. E {\bf 82}, 036120 (2010).

\bibitem{ravetti_2014}
M. G. Ravetti, L. C. Carpi, B. Goncalves, A. C. Frery and O. A. Rosso
PLoS ONE {\bf 9}, e108004 (2014).

\bibitem{bandt_2007} C. Bandt and F. Shiha,
J. Time Series Analysis {\bf 28}, 646-665 (2007).

\bibitem{aragoneses_2014} A. Aragoneses, S. Perrone, T. Sorrentino, M. C. Torrent, and C. Masoller,
Sci. Rep. {\bf 4}, 4696 (2014).

\bibitem{natcomm} D.V. Churkin, S. Sugavanam, N. Tarasov, S. Khorev, S.V. Smirnov, S.M. Kobtsev, S.K. Turitsyn
 Nature communications, {\bf 6} (2015).

\bibitem{jang2013ultraweak}  J.K. Jang, M. Erkintalo, S. Murdoch, and S. Coen,
 Nature Photonics {\bf 7}, 657-663 (2013).

\bibitem{garbin2015topological} B. Garbin, J. Javaloyes, G. Tissoni, and S. Barland,
 Nature communications, {\bf 6} (2015).

\end{thebibliography}
\end{document}